\documentclass[twocolumn]{aastex7}
\usepackage{amsmath}

\graphicspath{{./}{figures/}}

\begin{document}

\title{Formation of over-massive black holes in high-redshift disk galaxies via globular cluster accretion}

\author[orcid=0000-0002-1319-3433,sname='Yajima']{Hidenobu Yajima}
\affiliation{Center for Computational Sciences, University of Tsukuba, Ten-nodai, 1-1-1 Tsukuba, Ibaraki 305-8577, Japan}
\email[show]{yajima@ccs.tsukuba.ac.jp}

\begin{abstract}
Recent observations with the \textit{James Webb Space Telescope} (JWST) have suggested the existence of over-massive black holes (OMBHs) in high-redshift galaxies. 
In this paper, we propose a new mechanism for the formation of OMBHs, based on the accretion of globular clusters (GCs) in compact disk galaxies. 
We derive the conditions under which OMBHs can form, focusing on key parameters such as halo mass, redshift, and halo spin parameter. 
Our results show that at redshift $z = 10$, a halo with mass $10^{11}~M_{\odot}$ and a spin parameter of $\sim 0.02$ can form a black hole of $2.3 \times 10^{8}~M_{\odot}$ through GC migration and accretion via tidal disruption events (TDEs). 
The resulting black hole-to-stellar mass ratio can reach $\sim 0.1$, corresponding to the fraction of GC mass accreted onto the black hole. 
This mechanism thus provides a plausible explanation for the OMBHs observed by JWST. 
Furthermore, by combining our model with the halo mass function and the spin-parameter distribution, 
 we construct black hole mass functions that reproduce the number densities of the massive BH candidates UHZ1 and GHZ9 at $z \approx 10$, as well as the abundances of BHs with masses $\gtrsim 10^{8}\,\rm{M_\odot}$ at $z \approx 5$ inferred from JWST observations. However, our model overpredicts the abundance of BHs with masses $ < 10^{8}\,\rm{M_\odot}$, suggesting that moderately massive, inactive BHs are more frequent.
\end{abstract}


\section{Introduction} 

Understanding the formation of supermassive black holes (SMBHs) and their co-evolution with host galaxies is one of the major challenges in modern astrophysics. In the local universe, it is well established that SMBHs reside at the centers of galaxies and that their masses are tightly correlated with the stellar masses of the galactic bulges \citep[e.g.,][]{Ferrarase00, Gebhardt00, Merritt01, Gultekin09, Kormendy13, McConnell13, Reines15}. However, the physical mechanisms governing the formation of SMBHs and their co-evolution with galaxies remain poorly understood \citep{Volonteri10b, Inayoshi20}. 

In recent years, SMBHs with masses exceeding $10^9\,M_\odot$ have been discovered during the epoch of cosmic reionization \citep{Willott07, Mortlock11, Banados16, Fan19}. The exact mechanism by which black holes acquire enormous mass in a short period of time is not well understood, and it remains one of the great mysteries. Furthermore, wide-field surveys have identified a significant population of low-mass active galactic nuclei (AGNs) with masses in the range of $10^7$--$10^8\,M_\odot$ \citep{Matsuoka19, Izumi19}. In addition, observations with the James Webb Space Telescope (JWST) have revealed a large number of AGNs \citep{Harikane23b, Maiolino24}. In particular, a new population of AGNs, little red dots (LRDs), has been discovered \citep{Matthee24, Greene24}. 
These observations indicate that the abundance of SMBHs in the early universe may be significantly higher than previously anticipated.
One of the key features is the relationship between the black hole (BH) mass and the stellar mass, which is significantly larger compared to galaxies in the local Universe \citep{Maiolino24}. Here we call them over-massive black holes (OMBHs). Theoretically, the coevolution of black holes and galaxies is expected to correlate through feedback from AGN and stars \citep{DiMatteo05, Sijacki09, Sijacki15}. Recent cosmological galaxy formation simulations have shown that in low-mass halos of the early universe, supernova feedback (SN) reduces the gas density around black holes, thereby inhibiting their growth \citep{Dubois15, Rosas-Guevara16, Weinberger18, Zhu22}. Therefore, these observations suggest that there may be a rapid black hole growth mechanism that can not be followed in current cosmological simulations.

One possibility is the formation of massive seed black holes. Recent detailed simulations of Population III star formation suggest that the initial mass function shows a distribution ranging from 10 to 1000 solar masses \citep{Hosokawa11, Hosokawa16, Susa14, Hirano15, Stacy16, Sugimura20, Sugimura23, Latif22}. However, it is expected that strong ultraviolet radiation from nearby star-forming galaxies will cause the destruction of hydrogen molecules, leading to the formation of supermassive stars with masses $\sim 10^{5-6}~\rm M_{\odot}$ \citep[e.g.,][]{Omukai01, Bromm03, Regan09, Regan14, Inayoshi14a, Shlosman16, Latif16, Kimura25}.
These stars can form in primordial gas under strong Lyman-Werner radiation with $J_{21} \gtrsim 1000$ or higher \citep[e.g.,][]{Sugimura14, Dijkstra14, Agarwal14}.
Therefore, their formation sites are likely limited \citep{Chiaki23}. On the other hand, for Population II stars, in the case of very compact star clusters, the cluster can contract due to two-body relaxation processes, and massive stars migrate toward the center, resulting in formation of very massive stars with mass $\sim 1000~\rm M_{\odot}$ \citep{Portegies-Zwart04, Omukai08, Katz15, Kroupa20, Fujii24}. These could also serve as candidates for heavy seed black holes. However, for the collapse of the star cluster to occur on a timescale shorter than the lifetime of a massive star, special conditions such as a highly compact or top-heavy initial mass function (IMF) are required \citep{Yajima16a, Dekel25}.

Another scenario is the rapid growth of black holes (BHs) via super-Eddington accretion. Even when gas accretes spherically, if certain density and temperature conditions are met, radiation feedback becomes ineffective, and accretion proceeds at the Bondi accretion rate \citep{Park12, Inayoshi16, Toyouchi19}. Similarly, when gas accretes along a disk, there is the possibility that gas can accrete efficiently from the equatorial plane \citep{Sugimura17, Takeo18}. In this model, a major uncertainty lies in how, on a galactic scale, the black hole can accumulate large amounts of gas against the star formation and the stellar feedback. Additionally, as the galaxy evolves, heavy elements and dust are released into the interstellar medium by type-II supernovae. When interstellar dust is present, radiation pressure becomes significant, making gas accretion more difficult \citep{Yajima17b, Toyouchi19, Ogata21}.
Although these models could be potential candidates for SMBH formation, they require various stringent conditions, and it is unclear to what extent they can statistically explain the current number of SMBHs. 
In addition, it remains challenging to explain the OMBH formation based on super-Eddington accretion in the context of galaxy evolution. 
Recent simulations gradually connect the scales between a galaxy and an accretion disk around a BH \citep{Hopkins24, Shin25}. However, the period of galaxy evolution is quite limited due to the expensive calculations. Therefore, the co-evolution of a BH with a galaxy is still puzzling. 

In addition, stellar accretion may also contribute to the growth mechanism of black holes (BHs). When a star approaches sufficiently close to a BH, it can be tidally disrupted, with part of its gas accreting onto the BH \citep{Rees88}. This phenomenon is known as a tidal disruption event (TDE). Recent observations have detected TDE candidates from high-redshift galaxies \citep{Karmen25, DeCoursey25}. \citet{Yajima16a} semi-analytically estimated BH growth through TDEs, following the inward migration of massive stars via two-body relaxation in compact star clusters, the formation of very massive stars, and the subsequent direct collapse into BHs. They also suggested that the BHs grow via stellar accretion due to the resonant relaxation process. \citet{Kashiyama16} modeled star formation and stellar accretion onto BHs in the vicinity of galactic centers. More recently, \citet{Wang25} proposed the formation and accretion of Population III stars in galactic centers originating from cold streams. Furthermore, \citet{Dekel25} modeled the formation of high-density star clusters within galaxies, the formation of BH seeds therein, angular momentum transport and inward migration through interactions between these BHs, and their growth via mergers. While these models ideally describe the formation of massive black holes (MBHs), their co-evolution with galaxies and associated statistical properties remain poorly understood.

Here, we propose a new model for the rapid growth of BHs by compact star clusters, such as globular clusters (GCs). Recent spatially resolved emission line observations by JWST and ALMA have revealed that high-redshift galaxies retain compact galactic disks \citep{Tokuoka22, Xu24}. Additionally, JWST observations using gravitational lensing have discovered multiple compact star clusters within high-redshift galaxies \citep{Vanzella23, Adamo24, Fujimoto24}. Therefore, through the analysis of the age and metal content of globular clusters in nearby galaxies, the formation of globular clusters within galaxies during the expected epoch of cosmic reionization is now being observed.

Considering these conditions, it is possible that at high redshift, disk galaxies are formed and maintained, with multiple young globular clusters simultaneously. Additionally, emission line observations by JWST suggest that the gas density in star-forming regions is extremely high \citep{Isobe23}. Therefore, here we propose a black hole growth model based on high-density gas disks at high redshift. In this model, the following three conditions are considered to model the rapid growth of black holes: (1) the disk is maintained against supernova feedback, (2) globular clusters are formed within the disk, and (3) globular clusters move toward the center on short timescales.
In this study, we analytically derive the conditions under which OMBH formation occurs as a function of halo mass, redshift, and halo spin parameter. Furthermore, by combining these conditions with the halo mass function and the spin parameter distribution function, we study the statistical properties of OMBH formation and compare them with the recent JWST observational results.
Here, the cosmological parameters are considered as $H_{\rm 0}=67.4~{\rm km~s^{-1}~Mpc^{-1}},~\sigma_{8}=0.81, ~\Omega_{\rm m}=0.315, \Omega_{\rm b}=0.049, ~{\rm and}~ \Omega_{\rm \Lambda}=0.685$ \citep{Planck20}.

This paper is organized as follows. In Section 2, we describe the basic analytic models. In Section 3, we present the conditions to induce the rapid growth of BHs. In addition, we show the relation between BH and stellar masses, as well as the BH mass functions. In Section 4, we discuss caveats in my model. We summarize our findings in Section 5.

\begin{figure*}[ht!]
\plotone{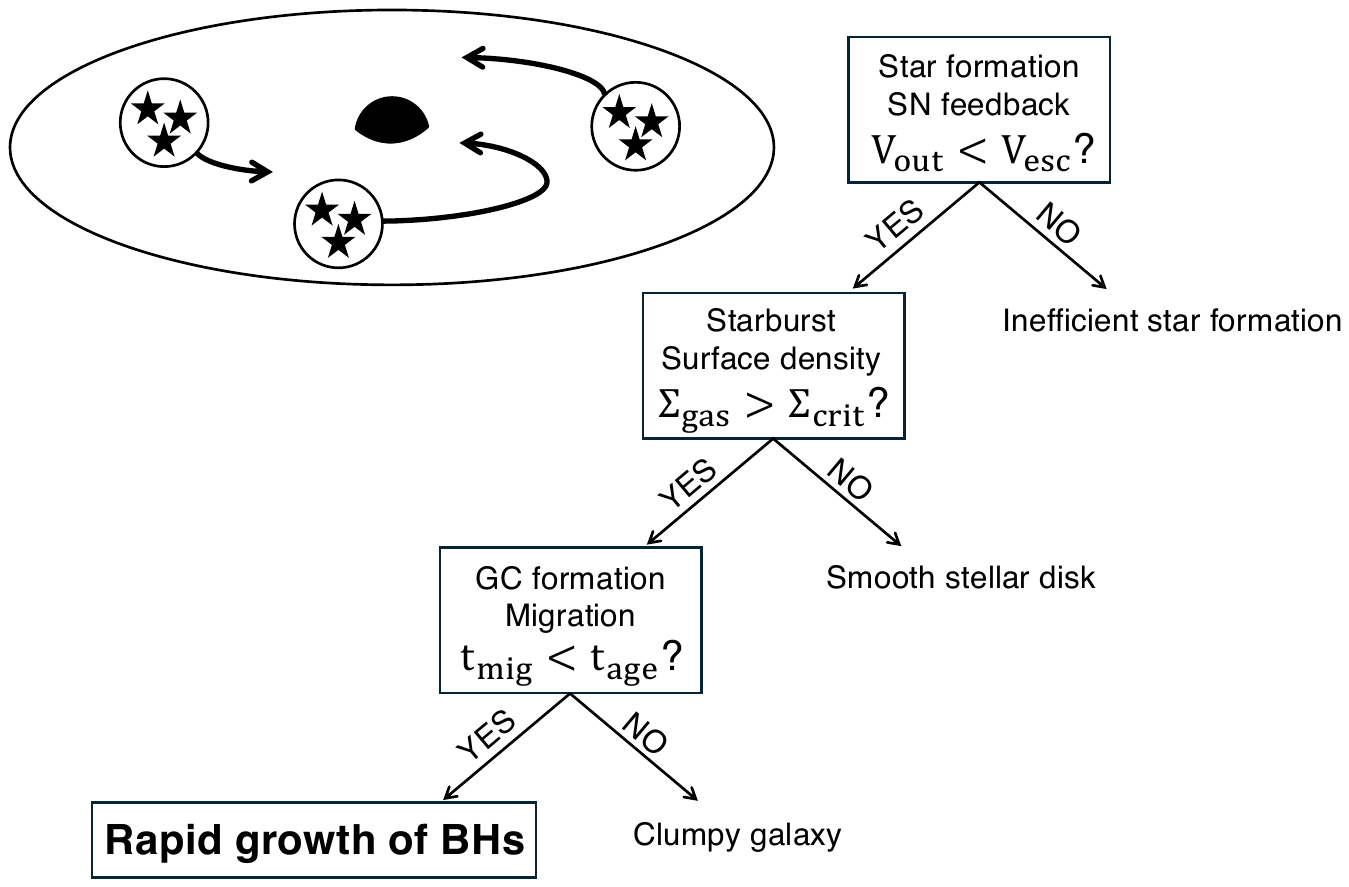}
\caption{Flow chart of the globular cluster accretion model for the rapid growth of black holes. If galaxies satisfy three conditions, the over-massive black holes form successfully. 
\label{fig:fig1}}
\end{figure*}

\section{Model} \label{sec:model}
We investigate the formation of star clusters in disk galaxies, their subsequent migration, and eventual accretion onto central black holes. Specifically, we consider the conditions under which gas is expelled from galaxies by supernova feedback, the gas surface density threshold required to form high-density clusters such as GCs, and the timescale constraints for cluster migration. These conditions are evaluated in terms of three key parameters: halo mass, redshift, and the halo spin parameter. We find that galaxies satisfying all of these criteria are capable of forming OMBHs. Figure 1 summarizes these conditions in the form of a flowchart.

Here, we consider an equilibrium state of an isothermal gas in the vertical direction with the pressure gradient force and the gravity force as
\begin{equation}
    \begin{split}
        c_{\rm s}^{2}\frac{d\rho}{dz} = - \rho \frac{d\phi}{dz} \\
        \frac{d^{2}\phi}{dz^{2}} = 4 \pi G \rho,
    \end{split}
\end{equation}
where $c_{\rm s}$ is the sound speed and $\phi$ is the gravitational potential.
Here, we consider the gravitational force exerted only by the gas.
By combining the above equations, we obtain an equation, 
\begin{equation}
\frac{d^2 ({\rm ln} \rho)}{dz^{2}}  = - \frac{4 \pi \rm G}{ c_{\rm s}^{2}} \rho.
\end{equation}
This gives a solution as \citep{Spitzer42} 
\begin{equation}
    \rho = \rho_{0} {\rm sech^{2}} \left( \frac{z}{z_{\rm 0}} \right),
\end{equation}
where $\rho_{0}$ is the density at the mid-plane of the disk.
By assuming the exponential distribution in the radial direction, we obtain a three-dimensional density structure of the gas as \citep[e.g.,][]{Spitzer42, Oh02}
\begin{equation}
\rho (r, z) = \rho_{0} ~ {\rm sech^2} \left(\frac{z}{z_{0}} \right) {\rm exp}\left( -\frac{2r}{R_{\rm d} }\right),
\end{equation}
where  $R_{\rm d}$ and $z_{0}$ are a scale radius and a vertical scale height, 
$z_{\rm 0} = c_{\rm s, eff} / (2 \pi G  \rho_{0}e^{-r/R_{\rm d}})^{1/2}$.
Here, we consider the effective sound speed $c_{\rm s, eff}$ instead of $c_{\rm s}$, which is defined as $c_{\rm s, eff} = \sqrt{c_{\rm s}^2 + \sigma_{\rm turb}^{2}}$, where $\sigma_{\rm turb}$ is the velocity dispersion due to turbulence fields.
In this work, we assume an isothermal warm gas with $10^{4}~\rm K$ and the mean molecular weight $\mu=1.22$, and $\sigma_{\rm turb} \approx c_{\rm s}$, corresponding to $c_{\rm s,eff} = 11.6~\rm km~s^{-1}$.
By assuming that the baryon conserves their specific angular momentum, $R_{\rm d}$ can be evaluated by $R_{\rm d} \sim \frac{\lambda}{\sqrt{2}} R_{\rm vir}$ \citep{Mo98}, where $\lambda$ is a halo spin parameter as $\lambda \equiv J \left|E\right|^{1/2}/GM^{5/2}$. 
Integration of the above profile is estimated as
\begin{equation}
\int \int 2 \pi r\rho(r,z) dz~dr = \left( \frac{8 \pi}{ {\rm G}}\right)^{1/2} c_{\rm s} \rho_{\rm 0}^{1/2} R_{\rm d}^{2}. 
\end{equation}
On the other hand, the disk mass is expressed as
\begin{equation}
    M_{\rm disk} = (1-\varepsilon_{\rm SF})f_{\rm d} \left( \frac{\Omega_{\rm b}}{\Omega_{\rm m}} \right) M_{\rm h},
\end{equation}
where $f_{\rm d}$ is the initial gas mass fraction of a galactic disk to the total gas mass in a halo and $M_{\rm h}$ is the halo mass. $\varepsilon_{\rm SF}$ is the star formation efficiency to the initial disk mass, i.e.,
\begin{equation}
    M_{\rm star} = \varepsilon_{\rm SF} f_{\rm d} \frac{\Omega_{\rm b}}{\Omega_{\rm m}} M_{\rm h}.
\end{equation}

At $z \gg 1$, the relation between $M_{\rm h}$ and $R_{\rm vir}$ is expressed as
\begin{equation}
    M_{\rm h} \approx \frac{H_{\rm 0}^2 \Omega_{\rm m, 0} \Delta_{\rm c}}{2 {\rm G}} R_{\rm vir}^{3} (1+z)^{3},
\end{equation}
where $\Delta_{\rm c}\approx 178$.
Combining Eq. (5), (6), and (8), we obtain $\rho_{0}$ as
\begin{equation}
\begin{split}
   \rho_{0} &= 1.8 \times 10^{-20}~{\rm g~cm^{-3}}  \left( \frac{f_{\rm d}}{0.2} \right)^{2} \left( 1-\varepsilon_{\rm SF}\right)^{2} \\
     &\times \left( \frac{T_{\rm gas}}{\rm 10^{4}~K}\right)^{-1}\left( \frac{M_{\rm h}}{10^{10}~\rm M_{\odot}} \right)^{2/3} \left( \frac{\lambda}{0.03} \right)^{-4} \left( \frac{1+z}{10} \right)^{4},
     \end{split}
\end{equation}
where $T_{\rm gas}$ is the gas temperature in a galactic disk.
%
By integrating the density profile of Eq. (1) along the vertical direction,  the surface density is estimated as
\begin{equation}
\begin{split}
\Sigma_{\rm gas} &= 2\rho_{0}z_{0}{\rm exp}\left( - \frac{2r}{R_{\rm d}}\right) \\
&=2.3 \times 10^{3}~{\rm M_{\odot}~pc^{-2}}~{\rm exp}\left( - \frac{r}{R_{\rm d}} \right) \left( \frac{M_{\rm h}}{10^{10}~\rm M_{\odot}} \right)^{1/3}  \\
&~~~~~~~~\times \left( 1-\epsilon_{\rm SF}\right) \left( \frac{f_{\rm d}}{0.2} \right) \left( \frac{\lambda}{0.03} \right)^{-2} \left( \frac{1+z}{10}\right)^{2}.
\end{split}
\label{eq:sigma}
\end{equation}

Next, we consider the formation of star clusters in the disk. The star formation in a molecular cloud proceeds with stellar feedback \citep{Dale12, Dale14, Krumholz12, Geen16, Kim18, Fukushima20}. 

A gas-rich disk can be unstable, resulting in fragmentation and cloud formation. The typical mass likely ranges $\sim$ Jeans mass to Toomre mass with a power-law function and the peak mass at the Jeans one \citep[e.g.,][]{Tasker09, Tasker11}. Therefore, we assume that cloud formation occurs with the Jeans mass, including the effect of turbulence described above. However, we regulate the cloud mass if its radius is larger than the disk scale height as $r_{\rm cl} = {\rm min} (r_{\rm J}, z_{\rm 0})$. Recent radiative-hydrodynamics simulations suggested that the SFE of a cloud sensitively depends on the initial physical properties of the cloud \citep{Kim18, Grudic21, Fukushima20, Fukushima21, Fukushima22, Fukushima23}. \citet{Fukushima21} derived the SFEs of clouds with three-dimensional radiative hydrodynamics simulations. They showed that the SFE depends on the surface density of the gas. 
We estimate it by
\begin{equation}
    \Sigma_{\rm cl} = \frac{m_{\rm cl}}{\pi r^{2}_{\rm cl}}.
\end{equation}
In relation to the galactic disk, we consider $ \Sigma_{\rm cl} \sim (r_{\rm cl}/z_{0}) \Sigma_{\rm gas}$.

We consider two modes of star formation in a cloud: one dominated by radiation pressure, and the other by photoionization. 
In cases of massive compact clouds, even ionized hot gas can keep collapsing due to a deep gravitational potential well. 
This happens if $\Sigma_{\rm cl} \gtrsim 300 ~\rm M_{\odot}~pc^{-2}$ \citep{Fukushima21}.
Then, the star formation can be suppressed when the radiation pressure balances with the gravitational force. 
The radiation pressure is estimated as
\begin{equation}
    F_{\rm rad} = \frac{L}{c} = \frac{\varepsilon_{\rm SF, cl} m_{\rm cl} l_{*} }{ c}
\end{equation}
where $\varepsilon_{\rm SF, cl}$ is the SFE to a cloud, $l_{*} = 1.3 \times 10^{3}~\rm L_{\odot}~M_{\odot}^{-1}$ is emissivity per stellar mass. 
It should be noted that if the UV optical depth of dust is not significantly larger than unity, $F_{\rm rad}$ is reduced by a factor of $(1 - e^{-\tau})$. In our case, however, because the system with a high gas column density is considered, the optical depth readily exceeds unity. 
Using a thin-shell approximation, the gravitational force from stars and gas is estimated by \citep{Kim16}
\begin{equation}
    F_{\rm grav} = \frac{ Gm_{\rm sh}(m_{\rm star} + m_{\rm sh}/2)}{r_{\rm cl}^2}
    = \frac{G m_{\rm cl}^{2}}{2 r_{\rm cl}^{2}} (1 - \varepsilon_{\rm SF, cl}^{2}),
\end{equation}
where $m_{\rm sh}$ is the mass of the gas shell and it is $m_{\rm sh} = m_{\rm cl} (1 - \varepsilon_{\rm SF, cl})$. For a self-gravitating gas shell, the factor of
$1/2$ appears in the expression for the binding energy.
By taking the balance $F_{\rm rad} = F_{\rm grav}$, we derive $\varepsilon_{\rm SF, cl}$ as
\begin{equation}
    \frac{\varepsilon_{\rm SF, cl}}{1-\varepsilon_{\rm SF, cl}^{2}} = \frac{\pi c G }{2 l_{*}}\Sigma_{\rm cl}.
\end{equation}
This is solved by the quadratic formula. Next, we consider the photoionization-dominated mode.
UV radiation from young massive stars can ionize the hydrogen of the cloud, resulting in the hot-ionized gas with the temperature $T_{\rm HII} \gtrsim 10^{4}~\rm K$. Due to the thermal pressure from the ionized gas, a cloud can be disrupted against its self-gravity. We consider the equation of motion of a spherical gas shell with the thermal pressure, the radiation pressure, and the gravity as
\begin{equation}
    \frac{d}{dt} (m_{\rm sh} \dot{r}_{\rm sh}) = 4 \pi r_{\rm sh}^{2} \rho_{\rm HII} c_{\rm s, HII}^{2} + \frac{L}{c} - \frac{G m_{\rm sh}(m_{\rm star} + m_{\rm sh}/2)}{r_{\rm sh}^{2}}.
\end{equation}
When the photoionization feedback is dominant, i.e., $\Sigma_{\rm cl} < 300~\rm M_{\odot}~pc^{-2}$, the above equation is simplified by ignoring the terms of the radiation pressure and gravity as follows.
\begin{equation}
    \frac{d r_{\rm sh}^{3}\dot{r}_{\rm sh}}{dt} = 3 \beta c_{\rm s, HII}^{2} r_{\rm sh}^{2}, 
\end{equation}
where $\beta= \frac{\mu m_{\rm p} \varepsilon^{1/2} }{1-\varepsilon} \left( \frac{S_{*}}{\rho_{0} \alpha_{\rm B}}\right)^{2}$.
Taking into account a self-similar solution, the above equation results in
$r_{\rm sh} = \beta^{1/2}c_{\rm sh, HII} ~t $ and $\dot{r}_{\rm sh} = \beta^{1/2} c_{\rm sh, HII}$. 
We assume that the star formation continues until the expanding shell reaches the radius of a cloud. The time scale is $t_{\rm exp} = R_{\rm cl} / \dot{r}_{\rm sh} = R_{\rm cl}/(\beta^{1/2} c_{\rm s, HII})$. Therefore, the stellar mass is estimated by
$m_{\rm star} = \varepsilon_{\rm ff} \left( \frac{t_{\rm exp}}{t_{\rm ff}} \right) m_{\rm cl}$, where $\varepsilon_{\rm ff}$ is the conversion rate from the gas to stars per free-fall time, which can be a few per cent \citep{Fukushima21}. 
Finally, the star formation efficiency is derived as 
\begin{equation}
    \varepsilon_{\rm SF} = \epsilon_{\rm ff} \beta^{-1/2} \left( \frac{t_{\rm cross}}{t_{\rm ff}}\right), 
\end{equation}
where $t_{\rm cross}=R_{\rm cl}/c_{\rm s, HII}$ is the sound crossing time. 
Thus, we summarize star formation efficiencies for the two modes as
\begin{equation}
\begin{split}
\varepsilon_{\rm SF, cl} = 
\begin{cases}
2.5 \times 10^{-2} \epsilon_{\rm ff,0.02}^{4/5}  T_{\rm HII,4}^{-14/25} \Sigma_{\rm cl, 100} \\
~~~~~~~~~~~~~~~~~~~~~~~~~~~~ {\rm for}~\Sigma_{\rm cl} < 300~\rm M_{\odot}~pc^{-2} \\
 0.5 \left( \sqrt{a^{2} \Sigma_{\rm cl, 100}^{-2} + 4} -  a \Sigma_{\rm cl, 100}^{-1} \right) \\
~~~~~~~~~~~~~~~~~~~~~~~~~~~~{\rm for}~\Sigma_{\rm cl} \ge 300~\rm M_{\odot}~pc^{-2},
\end{cases}
\end{split}
\end{equation}
where $a=\left( \frac{2 l_{\rm *}}{\pi \rm c G} \right) \left( 100~\rm M_{\rm \odot}~pc^{2}\right)^{-1} = 38.1$, $\varepsilon_{\rm ff, 002} \equiv \left( \frac{\varepsilon_{\rm ff}}{0.02}\right)$, $T_{\rm HII,4} \equiv \left( \frac{T_{\rm HII}}{10^{4}~\rm K} \right)$, and $\Sigma_{\rm cl, 100} \equiv \left( \frac{\Sigma_{\rm cl}}{100~\rm M_{\odot}~pc^{-2}}\right)$.

After a star cluster forms, member stars can move away from the birthplace and disperse because the gravitational potential becomes shallower due to the evaporation of the gas.
However, if $\varepsilon_{\rm SF, cl} \gtrsim 0.2-0.3$ \citep{Baumgardt07}, the stars can be bound due to their self-gravity. 
Here, we assume the critical surface density $\Sigma_{\rm crit} = 1000~\rm M_{\odot}~pc^{-2}$ as the formation condition of GCs. 
The critical surface density corresponds to $\varepsilon_{\rm SF, cl} = 0.25$.
Taking the assumption that $r_{\rm cl}=z_{0}$, $r=R_{\rm d}$, and $\varepsilon_{\rm SF}=0$, we derive the critical spin parameter $\lambda_{\rm crit}$ under which GCs can form as 
\begin{equation}
\lambda_{\rm crit} = 2.8 \times 10^{-2}  \left( \frac{f_{\rm d}}{0.2} \right)^{1/2} \left( \frac{M_{\rm h}}{10^{10}~\rm M_{\odot}} \right)^{1/6}   \left( \frac{1+z}{10}\right).
\end{equation}

\begin{figure}
\plotone{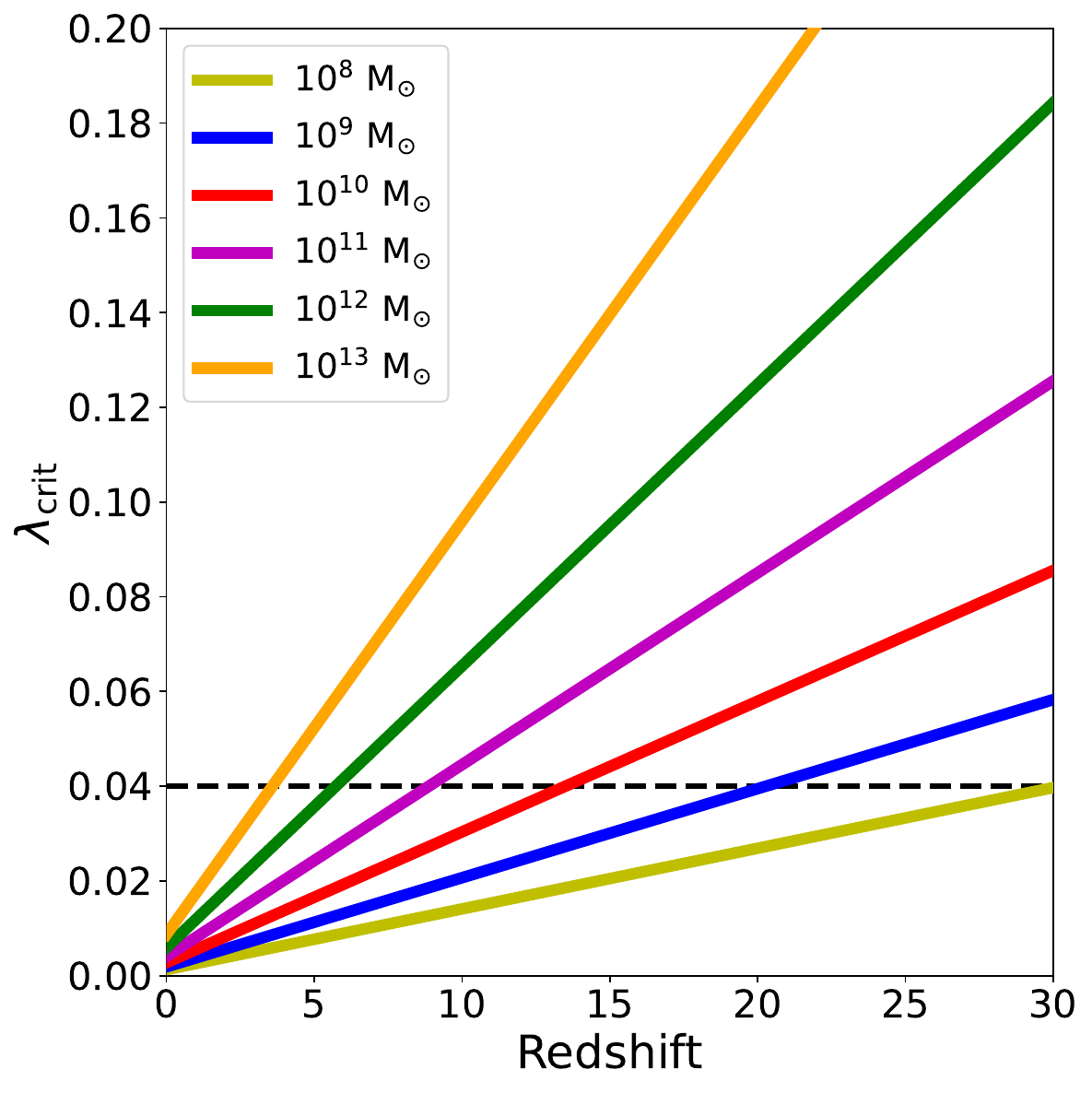}
\caption{Redshift evolution of the critical spin parameter under which globular clusters can form. Different colors of solid lines represent different halo masses. The black dashed line indicates the peak value of a halo spin parameter distribution \citep{Bullock01}.
}
\label{fig:lam_crit}
\end{figure}

Figure~\ref{fig:lam_crit} shows the redshift evolution of $\lambda_{\rm crit}$. Since the virial radius decreases as the redshift increases with a factor of $(1+z)$, $\lambda_{\rm crit}$ increases.
It is known that $\lambda$ shows a log-normal distribution with a peak value of $\lambda=0.04$ \citep{Bullock01}. 
In cases of halos with $M_{\rm h} = 10^{10}~\rm M_{\odot}$ and $\lambda =0.04$, the redshift for the GC formation is limited at $z \gtrsim 13$. 
Recent observations show multiple candidates for forming GCs in galaxies at $z \sim 6 -10$ \citep{Vanzella23, Adamo24, Fujimoto24}.
Our results suggest that the observed star clusters are formed in massive halos with $M_{\rm h} \gtrsim 10^{11}~\rm M_{\odot}$ or ones with rare low-spin parameters.

As the star formation proceeds, the gas surface density decreases. Then, even massive galaxies cannot form GCs when the SFE reaches a specific value. 
We derive the critical SFE for the GC formation ($\varepsilon_{\rm GC}$) from Eq. (\ref{eq:sigma}) as 
\begin{equation}
\begin{split}
\varepsilon_{\rm GC} = 1 &- 1.2 \left( \frac{\Sigma_{\rm crit}}{10^{3}~{\rm M_{\odot}~pc^{-2}}} \right) \left( \frac{r_{\rm cl}}{z_0}\right)^{-1} \left( \frac{f_{\rm d}}{0.2} \right)^{-1}  \\ ~
&\times \left( \frac{M_{\rm h}}{10^{10}~\rm M_{\odot}} \right)^{-1/3} \left( \frac{\lambda}{0.03} \right)^{2} \left( \frac{1+z}{10}\right)^{-2}.
\end{split}
\end{equation}
Here, we set $r=R_{\rm d}$ and assume that the cloud surface density is estimated as $\left( \frac{r_{\rm cl}}{z_{0}}\right) \Sigma_{\rm gas}$.

Globular clusters maintain their distance from the galactic center due to their initial angular momentum at formation. However, if there is a relative velocity between the clusters and the surrounding gas, dynamical friction from the gas is expected to reduce the angular momentum of the clusters, causing them to migrate toward the galactic center. The acceleration by the dynamical friction of the gas can be estimated as follows.
\begin{equation}
    a_{\rm DF} = -\frac{4 \pi {\rm G^{2}} M_{\rm cl} \rho_{\rm gas}}{v_{\rm rel}^{2}} I({\mathcal M}), 
\end{equation}
where $v_{\rm rel}$ is the relative velocity between a GC and the gas, $M_{\rm cl}$ is the mass of a star cluster,  $\rho_{\rm gas}$ is the background gas density, and
a factor $I(\mathcal M)$ is expressed by functions of Mach number as \citep{Ostriker99} 
\begin{equation}
I(\mathcal M) = 
\begin{cases}
     \dfrac{1}{2} {\rm ln} \left( \dfrac{1+ \mathcal M}{1 - \mathcal M}\right) - \mathcal M,  ~~{\rm for}~ {\mathcal M} < 1\\
     \dfrac{1}{2} {\rm ln} \left(1 - \dfrac{ 1}{\mathcal M^2}\right) + {\rm ln}~\Lambda,  ~~{\rm for}~ {\mathcal M} > 1,
\end{cases}
\end{equation}
where ${\rm ln}~\Lambda = {\rm ln}\left(r_{\rm max}/r_{\rm min}\right)$ is a term of Coulomb logarithm. 
The above functions are derived from a linear perturbation analysis. Note that \citet{Escala04} investigated the gaseous dynamical friction of BHs using hydrodynamics simulations and showed that $I(\mathcal M)$ should be modified for $\mathcal M \sim 1$. In this work, we consider star clusters moving with $\mathcal M \gg 1$ to the background gas disk. Therefore, using the above equation is reasonable. In general, it is difficult to estimate the Coulomb logarithm factor in non-uniform media.
\citet{Escala04} adopted ${\rm ln}~\Lambda= 3.1$.  \citet{Suzuguchi24} showed somewhat lower values of $I(\mathcal M)$ in the supersonic case than Ostriker's model due to a non-linear effect. Furthermore, the radiative feedback from a BH can reduce the drag force \citep{Park17, Sugimura20b, Toyouchi20, Ogata24}. Here, we assume a value ${\rm ln}~\Lambda=1.0$ considering a thin disk geometry.
Note that, in the case of dynamical friction on stars, the gas flow directly impacts a star and can thereby contribute to the drag force. Here, we do not take the direct impact into account.

Unlike stars and BHs, gas disks can maintain their structure not only through centrifugal force but also by balancing gravity with pressure from thermal motions and turbulence. Therefore, the rotational velocity of the gas is less than the Keplerian value. Consequently, stars and BHs in the galactic disk are expected to migrate inward to radii where gravitational and centrifugal forces balance, with orbital velocities greater than those of the gas. Recent observations have shown that the energy contribution from turbulence increases with redshift relative to rotational energy \citep{Wisnioski15, Neeleman20, Rizzo20, Parlanti23, deGraaff24, Xu24}. According to \citet{Tokuoka22}, a galaxy at redshift 9.1 shows $V_{\rm rot}/\sigma = 0.67$. On the other hand, galaxies dominated by rotational energy are also being discovered in the early universe \citep{Xu24}. Therefore, the relative velocity can range from the sound speed up to the galactic rotational velocity. In this study, we assume the relative velocity to be equal to the galactic rotational velocity as

\begin{equation}
\begin{split}
    v_{\rm c} &= \sqrt{\frac{GM_{\rm h}}{R_{\rm vir}}} \\
    &\approx 79.6~{\rm km~s^{-1}} \left( \frac{M_{\rm h}}{10^{10}~\rm M_{\rm \odot}}\right)^{1/3} \left( \frac{1+z}{10} \right)^{1/2}.
    \end{split}
\end{equation}

We estimate the migration time scale by 
\begin{equation}
\begin{split}
 t_{\rm mig}  \sim &v_{\rm c} / \left| a_{\rm DF} \right|\\
 \approx~ &1.3\times10^{2}~  {\rm Myr}~ \left( \frac{v_{\rm c}}{100~\rm km~s^{-1}} \right)^{3} \left( \frac{M_{\rm cl}}{10^{5}~\rm M_{\odot}}\right)^{-1}  \\
 &\times
\left( \frac{n_{\rm H}}{10^{4}~\rm cm^{-3}} \right)^{-1}  (1 - \varepsilon_{\rm SF})^{-1}
 I(\mathcal M)^{-1}
 \end{split}
\end{equation}
Here, we consider $\mathcal M = v_{\rm c} / c_{\rm s, eff}$.
The migration should occur within the cosmic age at a specific redshift, 
\begin{equation}
     t_{\rm age}(z) = \int_{z}^{\infty} \frac{dz'}{(1+z')H(z')}.
\end{equation}
\begin{figure}
\plotone{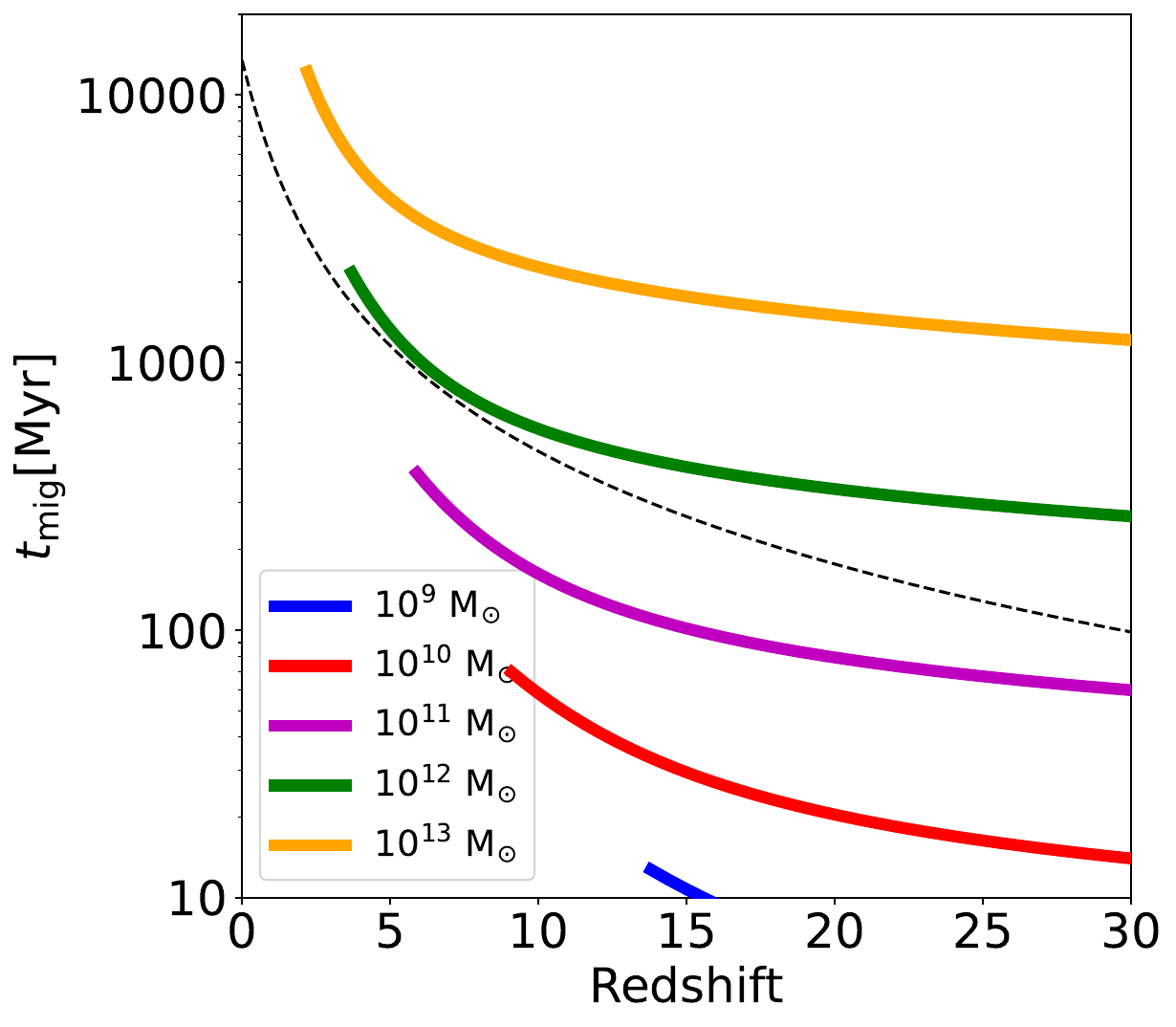}
\caption{Migration time of globular clusters into a galactic center as a function of redshift. Different colors of solid lines represent different halo masses with $\lambda=0.03$. The black dotted line shows the cosmic age at specific redshifts.}
\label{fig:tmig}
\end{figure}
Figure~\ref{fig:tmig} shows the migration time for hales with $\lambda=0.03$ and masses $M_{\rm h}=10^{9-13}~\rm M_{\odot}$ as a function of the redshift. Because of the higher circular velocity, the migration time in lower mass halos is shorter. At $z \approx 10$, massive halos with masses of $\gtrsim 10^{12}~\rm M_{\odot}$ show $t_{\rm mig}$ longer than the cosmic age. 
In cases of massive halos with $M_{\rm h} \gtrsim 10^{12}~\rm M_{\odot}$, $\lambda$ should be smaller than 0.03 because the gas density increases with a factor $\lambda^{-4}$ (see Eq. (2)), making $t_{\rm mig}$ shorter. 
Note that once a GC migrates near a BH via the above dynamical friction effect, other processes like type-I migration and two-body relaxation are likely to enhance the migration of a part of stars into a tidal radius \citep{Kashiyama16}. 

The migration time increases as the gas density decreases due to star formation and finally reaches the cosmic age. 
By taking the balance $t_{\rm mig} = t_{\rm age}$ with $r=R_{\rm d}$, we derive the critical star formation efficiency for the migration ($\varepsilon_{\rm mig}$) as 
\begin{equation}
\begin{split}
    \varepsilon_{\rm mig} = 1 - 0.26 \left( \frac{t_{\rm age}}{500~\rm Myr}\right)^{-1}
    \left( \frac{v_{\rm c}}{100~\rm km~s^{-1}} \right)^{3} \\
 \times \left( \frac{M_{\rm cl}}{10^{5}~\rm M_{\odot}}\right)^{-1}  
\left( \frac{n_{\rm H}}{10^{4}~\rm cm^{-3}} \right)^{-1}  I(\mathcal M)^{-1}.
\end{split}
\end{equation}

In the estimation of $\epsilon_{\rm{GC}}$ and $\epsilon_{\rm{mig}}$ in this study, we use $r = R_d$ for our calculations. In typical galactic disks, the inner regions are more unstable, and star formation progresses in an inside-out fashion. In such cases, stars in the inner regions move toward the center on shorter timescales. However, in an exponential disk, regions around $r \sim R_{\rm d}$ contribute significantly to the stellar mass. Therefore, in this study, we consider $r \sim R_{\rm d}$.

Then, we consider the condition for the SN feedback.
The SN feedback can reduce the gas density in a galactic disk as well as the star formation. Recent studies with cosmological simulations indicated that the supernova feedback induced a galactic wind that suppresses star formation, resulting in intermittent star formation histories in high-redshift low-mass galaxies \citep{Kimm14, Yajima17c, Yajima22, Yajima23, Sun23}. 
On the other hand, recent JWST observations suggest a high star formation efficiency in high-redshift galaxies \citep[e.g.,][]{Donnan23, Donnan24, Harikane23a, Harikane24}. This implies that, contrary to predictions from simulations, stellar feedback does not effectively suppress star formation in some galaxies. Moreover, if gas outflows are driven by feedback, it is generally difficult for galaxies to retain gas disks for long periods. Nevertheless, some high-redshift galaxies appear to host gas disks that are supported by rotation \citep{deGraaff24, Xu24}. Therefore, in this study, we estimate the conditions under which a halo-scale outflow driven by the SN feedback can occur in galaxies.
We consider the terminal momentum of an SN remnant when it enters the snow plow phase, which is estimated by \citep{Blondin98, Thornton98, Kim15, Kimm15}
\begin{equation}
    P_{\rm SN} \approx 3.0 \times 10^{5}~{\rm M_{\odot}~km~s^{-1}} ~E_{\rm SN,51}^{16/17} n_{\rm H}^{-2/17} \left( \frac{Z}{Z_{\odot}}\right)^{-0.14}, 
\end{equation}
where $E_{\rm SN, 51}$ is the energy injected by SNe normalized by $10^{51}~\rm erg$. Recent observations indicated that star-forming regions could be metal-enriched with $Z \gtrsim 0.1~Z_{\odot}$ \citep{Nakajima23, Bunker23, Isobe23b, Curti24}.
Therefore, we adopt $Z=0.1~\rm Z_{\odot}$.
The total momentum created in a galaxy is estimated as 
\begin{equation}
    P_{\rm SN, tot} = f_{\rm SN} M_{\rm star} P_{\rm SN}
\end{equation}
where $f_{\rm SN}$ is the frequency of SNe per unit stellar mass that is $\approx 1.2 \times 10^{-2} ~\rm M_{\odot}^{-1}$ for Chabrier initial mass function with a mass range of $0.1 - 100~\rm M_{\odot}$. Here we assume stars with $\ge 8~\rm M_{\odot}$ induce core-collapse supernovae.
Under the momentum conservation, the outflow velocity of the gas in a halo is estimated as
\begin{equation}
\begin{split}
    V_{\rm out} &= \frac{P_{\rm SN, tot}}{M_{\rm gas}} \\
    &= \frac{f_{\rm d}f_{\rm SN}P_{\rm SN} \varepsilon_{\rm SF}}{f_{\rm d}(1 - \varepsilon_{\rm SF}) + (1-f_{\rm d})}.
\end{split}
\end{equation}
If the outflow velocity exceeds the halo escape velocity, the SN feedback evacuates most of the gas from the halo, resulting in quenching of star formation. Therefore, we derive the critical SFE for the SN feedback ($\varepsilon_{\rm SN}$), taking the balance $V_{\rm out} = V_{\rm esc}$ as
\begin{equation}
    \varepsilon_{\rm SN} = \frac{V_{\rm esc}}{f_{\rm d}(f_{\rm SN}P_{\rm SN} + V_{\rm esc})}.
\end{equation}
Thus, we consider three types of critical SFEs for the GC formation, migration, and launching of galactic outflow. 
Taking the minimum of them, $\varepsilon_{\rm SF, min} = {\rm min}(\varepsilon_{\rm GC},  \varepsilon_{\rm mig}, \varepsilon_{\rm SN})$, we estimate the BH mass as
\begin{equation}
    M_{\rm BH} = f_{\rm TDE} \epsilon_{\rm SF, min} f_{\rm d}\frac{\Omega_{\rm b}}{\Omega_{\rm m}} M_{\rm h}, 
\end{equation}
where $f_{\rm TDE}$ is the mass fraction of a star that is accreted onto the black hole by TDE.
$f_{\rm TDE}$ changes with a collision impact factor and a mass fraction between the star and the BH, which is likely to be $\sim 0.1-0.5$ \citep{Ayal00, Guillochon13, Shiokawa15}. In this work, we assume $f_{\rm TDE} = 0.1$ as a default value. This assumes that most of the stars enter into their tidal radii. 
 Note that, however, $f_{\rm TDE}$ has a large uncertainty. Even if a stellar cluster undergoes tidal disruption, low-mass stars may be ejected because of two-body interactions between stars before being disrupted by a TDE. Therefore, we also consider the case with $f_{\rm TDE} = 0.01$, taking into account the effect of low-mass star ejection.
 The kicked low-mass stars then form a nuclear star cluster or a progenitor of the bulge \citep{Capuzzo-Dolcetta08, Antonini12, Arca-Sedda14}. Once an NSC or bulge is formed, stars begin to accrete onto the black hole on the loss cone filling timescale \citep{Merritt04, Merritt13, Vasiliev13}. This timescale can be longer than the galaxy’s dynamical time. Therefore, in this study, we consider only the stars from migrated clusters that directly enter the tidal radius as $f_{\rm TDE}$
, and we do not account for accretion due to loss cone filling through NSC or bulge formation. 

In this study, we assume that a seed BH with a mass comparable to or larger than that of a star cluster already exists at the galactic center prior to the star cluster migration process. The formation of such a seed BH could occur via a direct collapse of a supermassive star due to strong UV feedback  \citep{Regan09} or a very massive star, which forms when the collapse timescale of a star cluster is shorter than the lifetime of massive stars \citep{Portegies-Zwart04}. In addition, in high-density star clusters, stars can accrete onto the central BH through relaxation processes, further contributing to BH growth \citep{Yajima16a}. For simplicity, we consider only clouds with the Jeans mass in this work; however, when taking into account the distribution functions of clouds and star clusters, some clusters are expected to form seed BHs.

In order for a galaxy to keep a starburst, the SFE must be smaller than both $\varepsilon_{\rm GC}$ and $\varepsilon_{\rm SN}$, i.e.,
$\varepsilon_{\rm SF, min2} = {\rm min}(\varepsilon_{\rm GC}, \varepsilon_{\rm SN})$. We consider the mass ratio of BH to the stars when the starburst phase finishes as
\begin{equation}
    \dfrac{M_{\rm BH}}{M_{\rm star}} =
    \begin{cases}
    \dfrac{f_{\rm TDE}}{1-f_{\rm TDE}} 
    ~~~~~~~~~~~~~{\rm for~ \varepsilon_{\rm SF, min} = \varepsilon_{\rm GC}}~{\rm or}~ \varepsilon_{\rm SN}\\
    \dfrac{\varepsilon_{\rm SF, min}f_{\rm TDE}}{\varepsilon_{\rm SF, min}(1-f_{\rm TDE}) + (\varepsilon_{\rm SF, min2} - \epsilon_{\rm SF, min})} \\
    ~~~~~~~~~~~~~~~~~~~~~~~~~{\rm for~ \varepsilon_{\rm SF, min} = \varepsilon_{\rm mig}}.
    \end{cases}
\end{equation}
We implicitly assume that the gas not accreted onto the BH during a TDE is promptly used for subsequent star formation.

\section{Results} \label{sec:results}

\begin{figure}
\plotone{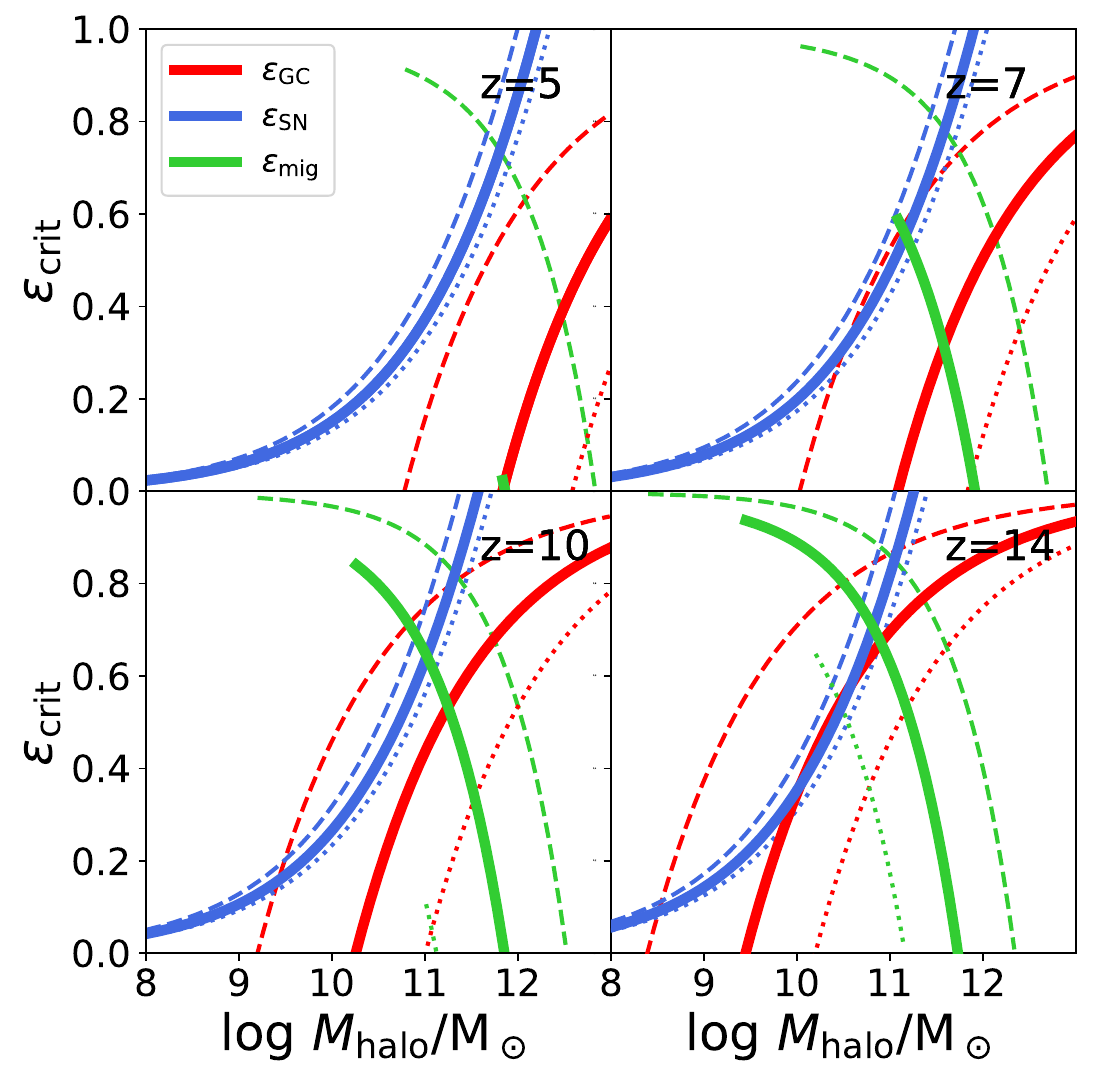}
\caption{Critical star formation efficiencies for the supernova feedback (blue lines), the globular cluster formation (red lines), and migration (green lines) as a function of halo mass. Solid lines show the cases with $\lambda = 0.03$. Dashed and dotted lines represent the cases with $\lambda = 0.02$ and $0.04$, respectively.}
\label{fig:esf_crit}
\end{figure}

As derived in the previous section, the formation of OMBHs requires that three conditions be satisfied, each associated with a limiting SFE. These limits are functions of halo mass, spin parameter, and redshift. Here, we analyze the dependence of the critical SFEs on these parameters. Figure~\ref{fig:esf_crit} shows the variation of the critical SFEs with halo mass at redshifts 5, 7, 10, and 14.

$\varepsilon_{\rm GC}$ increases with the halo mass as shown in Eq. (4). For example, a halo with $M_{\rm h} = 10^{11}~\rm M_{\odot}$ at $z=10$ can continue the formation of GCs even if more than half the disk mass is consumed. Low-mass halos at lower redshifts are likely to have unphysical negative values, which means that their initial surface densities do not satisfy the condition of the GC formation. 
As shown in Eq. (2), the surface gas density is highly sensitive to the spin parameter. If the spin parameter is smaller than 0.02, globular clusters (GCs) can form over a wide range of halo masses. On the other hand, such a low spin parameter requires an isotropic distribution of matter in the large-scale structure, making it statistically rare.

$\varepsilon_{\rm SN}$ also increases with the halo mass because the gravitational potential increases. 
Once the halo mass exceeds $\sim 10^{10-11}~\rm M_{\odot}$, it becomes evident that expelling the halo gas is difficult, even if most of the disk gas is converted into stars. This trend is also suggested by the stellar-to-halo mass ratio observed in the local universe \citep{Moster13, Behroozi13}. Furthermore, recent galaxy formation simulations have shown that in low-mass halos, the SN feedback can blow out the gas, leading to bursty star formation and reduced SFE.
As the redshift increases, the gravitational potential becomes deeper even for halos of the same mass, resulting in a higher value of $\varepsilon_{\rm SN}$. Regarding the spin parameter, although the density in the star-forming region becomes higher, 
$\varepsilon_{\rm SN}$ shows little dependence on the spin parameter because, as shown in Equation (21), the terminal momentum depends on the gas density insensitively.

Note that both GCs and SNe increase with halo mass in a similar manner, but GCs are generally smaller in magnitude. Therefore, after GC formation has ceased, star formation with low efficiency may proceed in molecular clouds. These stellar associations are likely to be disrupted after the molecular clouds evaporate, resulting in the formation of a smooth stellar disk. This suggests that a system in which GCs coexist with a smooth stellar disk may form. In contrast, at $z=14$, halos with masses of $\sim 10^{10-11}~\rm M_{\odot}$ have $\varepsilon_{\rm SN}$ lower than $\varepsilon_{\rm GC}$. In this case, GC formation may proceed while the system is depleted of gas by SN feedback, resulting in clumpy galaxies.

$\varepsilon_{\rm mig}$ shows the opposite trend to $\varepsilon_{\rm GC}$ and $\varepsilon_{\rm SN}$.
As the halo mass increases, the rotational velocity of the star cluster becomes larger, resulting in weaker dynamical friction from the gas and thus a longer migration time. Although the gas density also increases with halo mass, which enhances the dynamical friction for a star cluster with the same relative velocity, the dependence on halo mass is steeper (proportional to the 3/2 power). Therefore, $\varepsilon_{\rm mig}$ decreases with increasing halo mass. As shown in Figure 3, $\varepsilon_{\rm mig}$ reaches zero around $M_{\rm h} \sim 10^{12}~\rm M_{\odot}$ above which GCs cannot migrate to the galactic centers within a cosmic age, even for the initial gas-rich state. For the same halo mass, $\varepsilon_{\rm mig}$ sensitively depends on the spin parameter because the drag force is proportional to $\rho$, which strongly depends on the spin parameter.

\begin{figure}
\plotone{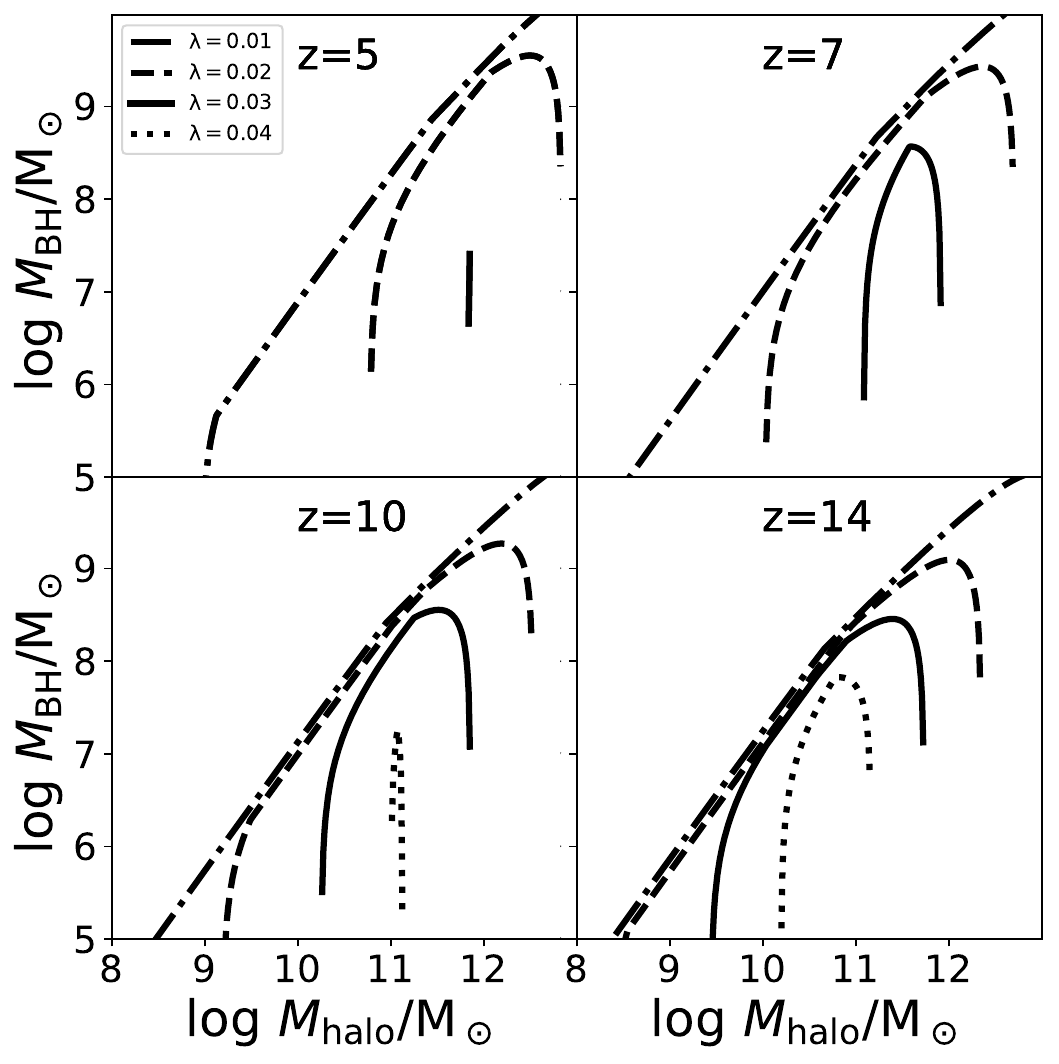}
\caption{Black hole mass as a function of halo mass. Different line types represent different spin parameters: dot-dashed for $\lambda=0.01$, dashed for $\lambda=0.02$, solid for $\lambda=0.03$, and dotted for $\lambda=0.04$.}
\label{fig:mbh_mhalo}
\end{figure}

Figure~\ref{fig:mbh_mhalo} illustrates the relationship between black hole mass and halo mass. Generally, up to a certain halo mass, the black hole mass increases with increasing halo mass. However, beyond a certain halo mass, as shown in Figure~\ref{fig:esf_crit}, black hole growth becomes inefficient due to the limitation imposed by the migration time. Consequently, for each spin parameter and redshift, a characteristic halo mass exists at which black hole growth is maximized, as well as a corresponding maximum black hole mass. For instance, at redshift $z=10$, when the spin parameter is 0.03, black holes can grow to $\gtrsim 10^{8}~\rm M_{\odot}$ within halos in the mass range of $8 \times 10^{10}$--$7 \times 10^{11}~\rm M_{\odot}$. Furthermore, when the spin parameter is lower than 0.02, our results suggest that black holes more massive than $10^{9} ~\rm M_{\odot}$ can still form even at $z=10$.

\begin{figure}
\plotone{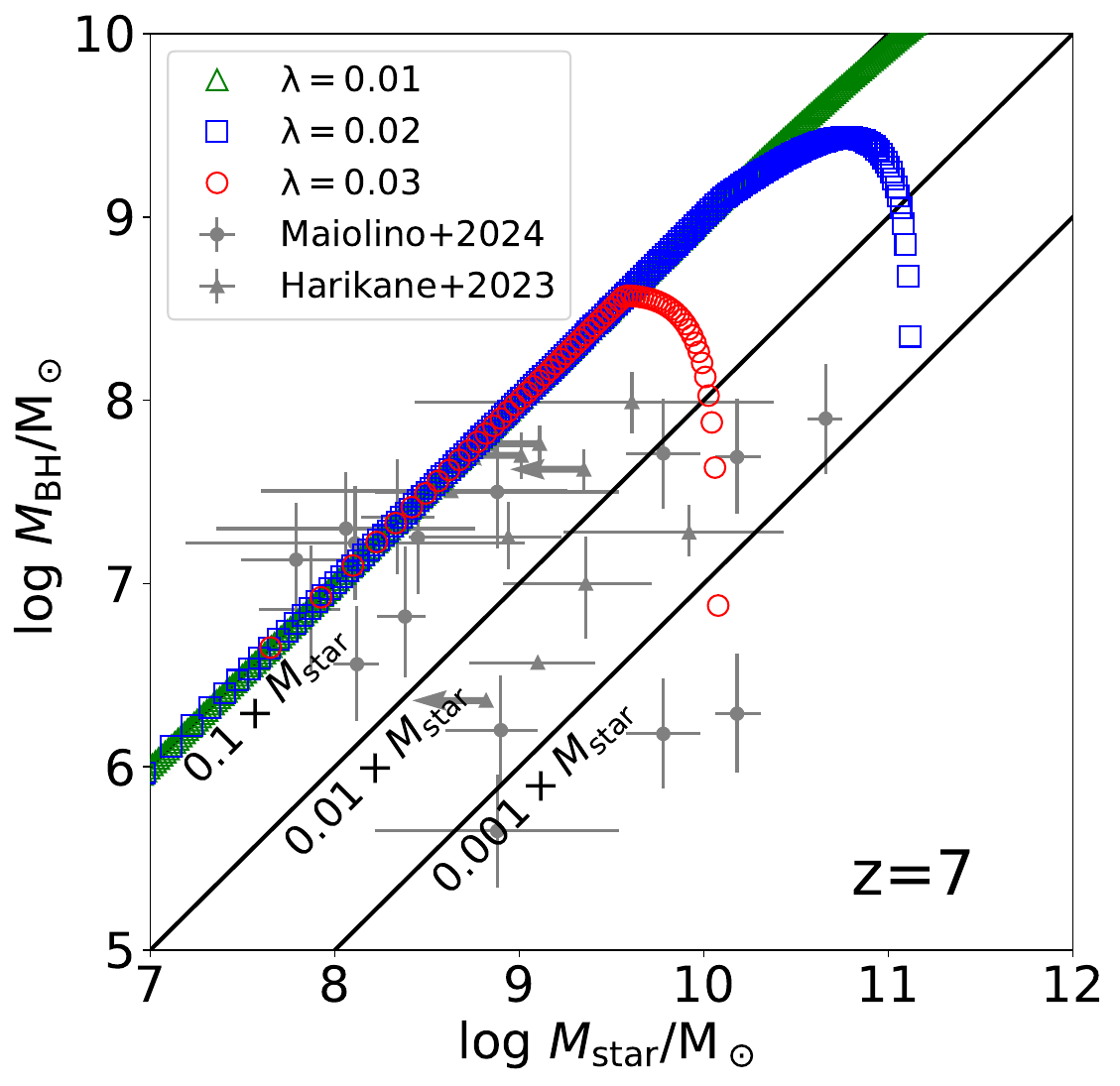}
\caption{BH mass as a function of stellar mass. Different open symbols represent results for different spin parameter cases: green triangles for $\lambda=0.01$, blue squares for $\lambda=0.02$, and red circles for $\lambda=0.03$. Filled circles and squares with error bars show observational results by \citet{Maiolino24} and \citet{Harikane23b}, respectively.}
\label{fig:mbh_mstar}
\end{figure}

Figure~\ref{fig:mbh_mstar} shows the relationship between black hole mass and stellar mass. Recent JWST observations suggest the existence of OMBHs with masses up to $\approx 10 ~\%$ of the stellar mass. Our model naturally reproduces such OMBHs. Below a certain halo mass, globular clusters efficiently migrate toward the center and, through TDEs, result in efficient gas accretion onto the black hole. As a result, the black hole to stellar mass ratio becomes constant at $f_{\rm TDE}$, potentially reaching the high values observed by JWST.

The BH mass exhibits a peak at a certain stellar mass and decreases beyond that point. Systems with smaller spin parameters continue to experience GC accretion onto the BH even when the SFE becomes large, resulting in high BH-to-stellar mass ratios even in massive systems. The stellar mass at the peak is $M_{\rm star} = 3.9 \times 10^{9}~\rm M_{\odot}$ for $\lambda = 0.03$ and $M_{\rm star} = 6.0 \times 10^{10}~\rm M_{\odot}$ for $\lambda = 0.02$.
Note that we ignore the evolutionary history of halos in this analysis. In reality, massive halos may have formed massive BHs during earlier epochs when their progenitor halos were less massive, and such BHs could already reside at the center. Therefore, the BH masses shown here should be regarded as lower limits, and massive galaxies may host even more massive BHs.

\citet{Behroozi13} successfully modeled the relationship between halo mass and stellar mass through a phenomenological approach that accurately reproduces the observed stellar mass function. In our model, stellar mass depends on both the halo mass and the spin parameter. However, we have verified that our model is able to approximately reproduce the observed stellar mass function at $z = 7$. Further statistical modeling of star formation and stellar mass will be addressed in future studies.

From the above analysis, it was found that the BH mass depends on the halo mass, redshift, and spin parameter. Next, we derive the BH mass function by using the halo mass function and the spin parameter distribution function. The halo mass function is calculated based on \citet{Sheth01}. The spin parameter distribution function is known to follow a log-normal distribution, as described below.
\begin{equation}
    P(\lambda) = \frac{1}{\sqrt{2\pi} \sigma \lambda}
    {\rm exp}\left( - \frac{{\rm ln^{2}}(\lambda/\lambda_{0})}{2 \sigma^{2}}\right),
\end{equation}
where we set the parameters as $\lambda_{0}=0.04$ and $\sigma=0.5$ \citep{Bullock01}.

\begin{figure*}
\plotone{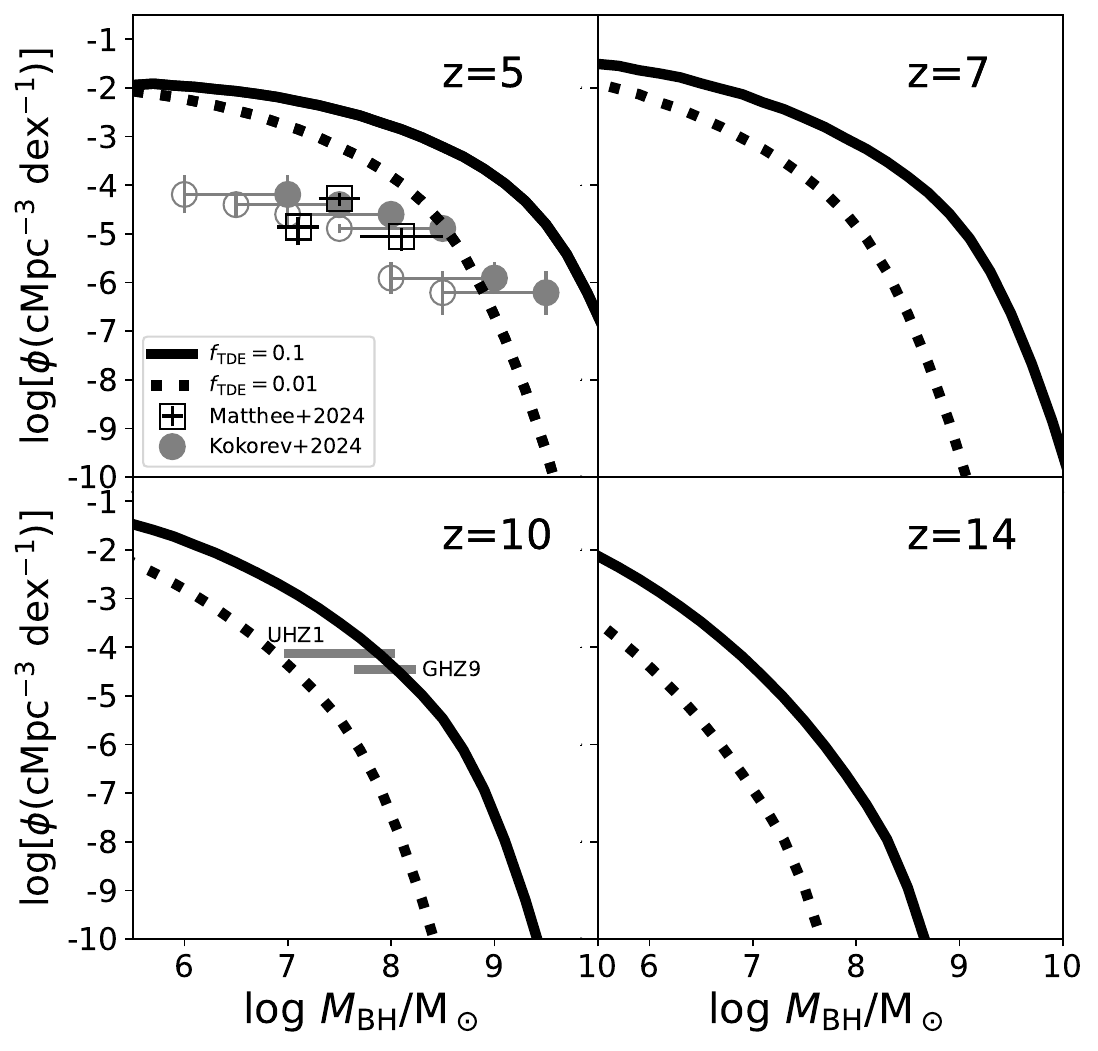}
\caption{BH mass functions. Solid and dotted lines represent the results with $f_{\rm TDE}=0.1$ and $0.01$, respectively. Open squares and filled circles $z=5$ show the observational results by \citet{Matthee24} and \citet{Kokorev24}, respectively.
UHZ1 and GHZ9 in the panel of $z=10$ are from observations \citep{Bogdan24, Fujimoto24b, Napolitano24}, summarized in \citet{Jeon25}.}
\label{fig:mbh_func}
\end{figure*}

Figure~\ref{fig:mbh_func} shows the BH mass function at each redshift.
Recent observations suggested the BH mass functions at $z \approx 5$ \citep{Matthee24, Kokorev24}. Our results successfully reproduce the observed number density of BHs with masses of $10^{8-9}~\rm M_{\odot}$. Furthermore, our model predicts a significantly higher number density of lower-mass BHs. In our model, we do not impose any constraints on the AGN lifetime or duty cycle. Therefore, even more black holes may exist than are observed. However, as discussed in the next section, various uncertainties may lead to either an overestimation or an underestimation of the number of massive black holes. In particular, we do not account for gas accretion onto BHs. Cosmological simulations of galaxy formation suggest that gas can accrete efficiently onto BHs, overcoming SN feedback in massive halos with $M_{\rm halo} \gtrsim 10^{11-12}~M_{\odot}$ \citep{Rosas-Guevara16, Dubois21, Yajima22}. For example, for a spin parameter of $\lambda = 0.02$, halos of $10^{12}~M_{\odot}$ can host BHs with masses of $\sim 2\times 10^{9}~M_{\odot}$. Therefore, such massive BHs may grow further, extending the high-mass end of the mass function to even larger values.

The BH mass function decreases with increasing redshift. This is because, for a given halo mass, the number density of halos becomes smaller at higher redshift, and also because the rotational velocities of galaxies become larger, increasing the relative velocities of BHs and thereby reducing the efficiency of their migration toward the galactic center. 
Recent observations have successfully detected AGNs at $z \approx 10$, GNZ-11, UHZ1 and GHZ9 \citep{Bunker23, Kovacs24, Bogdan24, Napolitano24}.
In particular, UHZ1 and GHZ9 indicate the high BH masses with $\sim 10^{7}-10^{8}~\rm M_{\odot}$ and the high number densities with $\sim 10^{-4}-10^{-5} ~\rm cMpc^{-3}~dex^{-1}$.
Our models nicely reproduce the mass and density. 
Furthermore, our model predicts SMBHs with $\sim 10^{8}~\rm M_{\odot}$ can exist even at $z \sim 14$, although its number density is low, $\sim 10^{-8}~\rm cMpc^{-3}$ .

\section{Discussion} \label{subsec:discussion}

In this study, we consider only supernova explosions as drivers of galactic outflows, while radiative feedback is primarily accounted for in the formation of star clusters within molecular clouds. However, in high-density gas environments, radiation pressure may also contribute significantly to the launching of galactic outflows. Here, we estimate and discuss the effects of radiative feedback.
In a gaseous disk, radiation emitted by stars formed near the mid-plane propagates toward both the upper and lower surfaces of the disk. During this process, dust absorbs the stellar radiation and re-emits it as infrared thermal radiation. If this infrared radiation is subsequently absorbed on the opposite side of the disk, the radiation pressure can be enhanced. Taking into account the re-absorption of infrared radiation, the radiative force can be estimated as follows.
\begin{equation}
\begin{split}
    F_{\rm rad, gal} &= \frac{L}{c}(1+\tau_{\rm IR}) \\
    &= \frac{\varepsilon_{\rm SF}l_{*}f_{\rm d}(\Omega_{\rm b}/\Omega_{\rm m})M_{\rm h}}{c}\left(1+\kappa_{\rm IR}\Sigma_{\rm gas} \right)
  \end{split}
\end{equation}
where $\tau_{\rm IR}$ is the optical depth of dust at the infrared wavelengths, $\kappa_{\rm IR}$ is the absorption coefficient for dust, 
\begin{equation}
\begin{split}
    \kappa_{\rm IR} = 2.5&~{\rm cm^{2}~g^{-1}}~
    \left( \frac{\bar{Q}_{\rm IR}}{10^{-2}}\right)
    \left( \frac{\bar{a}}{\rm 0.1~\mu m}\right)^{-1} \\
   & \times \left( \frac{\rho_{\rm d}}{3~\rm g~cm^{-3}}\right)^{-1}
    \left( \frac{D}{10^{-2}}\right)
    \left( \frac{Z}{\rm Z_{\odot}}\right),
\end{split}
\end{equation}
where $\bar{Q}_{\rm IR}$ is an absorption efficiency to the geometrical cross section, $\bar{a}$ is the radius of the dust grains, $\rho_{\rm d}$ is the mass density, and $D$ is the dust-to-gas mass ratio when the gas metallicity is at the solar abundance $Z=Z_{\odot}$. Here, we assume the dust mass increases with the metallicity \citep[e.g.,][]{Draine07}. 
The radiation force likely acts over the vertical scale of the disk, and the corresponding kinetic energy of the disk, neglecting the effect of gravity, is estimated as
\begin{equation}
E_{\rm rad} = F_{\rm rad, gal} z_{\rm d}.
\end{equation}
Beyond the disk scale, a gas shell forms where the opacity becomes negligible. We then compare this kinetic energy with that generated by the SN feedback. 
By assuming that the total momentum is uniformly distributed in the disk, the total kinetic energy due to the SN feedback can be estimated as
\begin{equation}
    \begin{split}
        E_{\rm SN} &= \frac{1}{2} \frac{P_{\rm SN,tot}^{2}}{f_{\rm d}(\Omega_{\rm b}/\Omega_{\rm m})M_{\rm h}} \\
        &=\frac{1}{2} f^{2}_{\rm SN} \varepsilon_{\rm SF}^{2} f_{\rm d}\left(\frac{\Omega_{\rm b}}{\Omega_{\rm m}}\right)M_{\rm h} P_{\rm SN}^{2}.
    \end{split}
\end{equation}
Thus, the energy ratio is derived by
\begin{equation}
\begin{split}
    \frac{E_{\rm rad}}{E_{\rm SN}} &= 
    \frac{2 l_{*}M_{\rm F}t_{\rm light,z0}}{f_{\rm SN}^{2}\varepsilon_{\rm SF}P_{\rm SN}^{2}}\\
    &=5.6 \times 10^{-2} \left( \frac{M_{\rm F}}{10}\right) \left( \frac{t_{\rm light, z0}}{100~\rm yr} \right) \left( \frac{\varepsilon_{\rm SF}}{0.1} \right)^{-1}
\end{split}
\end{equation}
where \( M_{\rm F} = \left(1 + \kappa_{\rm IR} \Sigma_{\rm gas} \right) \), and \( t_{\rm light,z0} \) is the light-crossing time defined as \( z_{0}/c \).
We assume that $n_{\rm H}=10^{4}~{\rm cm^{-3}}$ and $Z=0.1~\rm Z_{\odot}$.
The factor \( M_{\rm F} \) does not significantly exceed 10 in the low-metallicity environments. 
Thus, in most cases of high-redshift dwarf galaxies, the SN feedback dominates over radiation pressure. 
Therefore, the suppression of star formation by radiation force can be regarded as a secondary effect. 
On the other hand, recent theoretical models suggest that radiation pressure may be capable of driving dusty gas outflows, potentially reproducing the observed UV luminosity functions \citep{Ferrara25}.

%
%
%
%

In this work, we assume that a part of the GC mass accretes onto a BH if it migrates to a galactic center within a cosmic time. We also assume the presence of a seed BH at the center of a galaxy whose mass is comparable to or larger than that of a star cluster.
 However, if a BH mass is not much heavier than a GC, the BH can be kicked out due to 2-body interaction with the GC. \citet{Dekel25} showed that in the case of multiple MBHs, the BH can grow stochastically, and once the BH becomes much heavier than the others, it can grow stably thereafter. 

In this study, we have analytically estimated the formation of OMBHs based on the idealized model. In reality, the cosmological evolution of galaxies involves various processes, including galaxy mergers, gas accretion along filaments, and global instabilities in galactic disks. In addition, gas inflow from galaxies toward the central BH is also a crucial process for BH growth \citep{Dubois15, Yajima17b, Inayoshi22b, Shin25}. 
If gas accretion proceeds efficiently, the black hole (BH) mass can grow even larger. However, recent cosmological simulations have shown that the gas near the galactic center is depleted by supernova and AGN feedback, which can easily hinder BH growth \citep{Rosas-Guevara16, Dubois21}. Therefore, the formation of OMBHs may be difficult if only gas accretion is considered. On the other hand, the efficiency of feedback depends on the gas density and clumpiness, so assessing how much gas accretion onto the BH is suppressed requires higher-resolution simulations.
Therefore, these processes remain challenging to model accurately, even in recent high-resolution galaxy simulations. In future work, we aim to model the co-evolution of galaxies and black holes more realistically by performing cosmological hydrodynamics simulations.

Our model suggests that TDEs frequently occur in a short period once the halo satisfies the three conditions. A part of the gas accretes onto a BH, likely involving radiation from an accretion disk and high-energy transients associated with a jet. However, so far, variability has not been observed for most LRDs \citep{Kokubo24, Tee25, Furtak25}, although the number of observed epochs is restricted. Our model suggests that a large number of stars may accrete onto BHs via TDEs on a timescale comparable to the dynamical time of their host halos. In such cases, the TDE frequency could be shorter than a few years. Therefore, it is possible that the variability is not detected simply because the frequency is too high. Moreover, if the BH mass exceeds $\sim 10^{8}~\rm M_{\odot}$, the tidal radius can be smaller than a horizon scale. Therefore, observable signatures can be suppressed in the case of SMBHs that are hosted by massive halos with a low spin parameter.

\section{Summary} \label{sec:summary}

In this paper, we have modeled the growth of BHs within compact disk galaxies. In particular, we proposed a new scenario for the formation of OMBHs in which GCs formed in the disk migrate toward the galactic center, and a fraction of their mass accretes onto the BH, potentially leading to rapid BH growth. For this process to occur, we have identified three necessary conditions: (1) the absence of large-scale galactic outflows driven by supernova feedback, (2) sufficiently high gas surface density in the disk to enable the GC formation, and (3) a GC migration timescale shorter than the age of the universe. Based on these criteria, we have derived constraints on halo mass, spin parameter, and redshift.

Our results show that at redshift $z \sim 10$, a halo with mass $10^{11}~M_{\odot}$ and a spin parameter of $\sim 0.02$ can form a black hole of $2.3 \times 10^{8}~M_{\odot}$ through GC migration and accretion via TDEs. 
The resulting black hole-to-stellar mass ratio can reach $\sim 0.1$, corresponding to the fraction of GC mass accreted onto the black hole. 
This mechanism thus provides a plausible explanation for the OMBHs observed by JWST. 

Recent observations with JWST have revealed a large number of massive BH candidates. Their BH-to-stellar mass ratios are estimated to be 10 to 100 times higher than those observed in local galaxies, reaching values as high as $\sim 0.1$ \citep{Maiolino24}. Our model suggests that such high mass ratios can be explained by the mass fraction of stellar clusters that are disrupted via TDEs and subsequently accreted onto the central BH. Furthermore, by combining our model with the halo mass function and the spin parameter distribution function, we constructed BH mass functions at various redshifts. We found that our model naturally reproduces the number density of SMBHs reported in recent JWST observations.
In addition, our models predict that SMBHs with a mass of $10^{8-9}~\rm M_{\odot}$ can exist in a cosmic volume with $(100~\rm cMpc)^{3}$ at $z=10$ and $(1~\rm cGpc)^{3}$ at $z=14$.

The BH mass function predicted by our model is sufficiently larger than estimates from cosmological simulations. However, current galaxy simulations face significant challenges in simultaneously incorporating galaxy-scale dynamics, globular cluster formation, their migration toward the galactic center, and mass accretion onto BHs via TDEs. As a result, previous galaxy simulations may have underestimated BH masses by overlooking the processes proposed in this study. In future simulation studies, we aim to incorporate these mechanisms to construct a more realistic model for the co-evolution of BHs and their host galaxies.


\begin{acknowledgments}
We wish to thank the anonymous referee for detailed comments and suggestions that improved this paper. 
We thank Kazumi Kashiyama, Hajime Fukushima, Kohei Inayoshi, Kazuyuki Omukai, and Masayuki Umemura for fruitful discussions. 
The numerical simulations were performed on the computer cluster, XC50 and XD2000 in NAOJ, and Trinity at Center for Computational Sciences in University of Tsukuba. This work is supported in part by MEXT/JSPS KAKENHI Grant Number 21H04489, and JST FOREST Program, Grant Number JP-MJFR202Z. 
For the purpose of open access, the author has applied a Creative Commons Attribution (CC BY) licence to any Author Accepted Manuscript version arising from this submission.
\end{acknowledgments}

%




\bibliography{HY}{}
\bibliographystyle{aasjournalv7}



\end{document}